# Nanoscale electronic transport at graphene/pentacene van der Waals interface.

Michel Daher Mansour[1], Jacopo Oswald[2,3], Davide Beretta[2], Michael Stiefel[2,§], Roman Furrer[2], Michel Calame[2,3,4,*], Dominique Vuillaume[1,*].

1. *Institut for Electronic, Microelectronic and Nanotechnology, Centre National de la Recherche Scientifique, Villeneuve d'Ascq, France.*
2. *Transport at Nanoscale Interfaces Laboratory, Swiss Federal Laboratories for Material Science and Technology (EMPA), Dübendorf, Switzerland.*
3. *Swiss Nanoscience Institute, University of Basel, Switzerland.*
4. *Department of Physics, University of Basel, Switzerland*

§ Current address : IBM, Binnig and Rohrer Nanotechnology Center, Zurich, Switzerland.
* Corresponding authors: dominique.vuillaume@iemn.fr; michel.calame@empa.ch

**Abstract.**

We report a study on the relationship between structure and electron transport properties of nanoscale graphene/pentacene interfaces. We fabricated graphene/pentacene interfaces from 10-30 nm thick needle-like pentacene nanostructures down to two-three layers (2L-3L) dendritic pentacene islands, and we measured their electron transport properties by conductive atomic force microscopy (C-AFM). The energy barrier at the interfaces, *i.e.* the energy position of the pentacene highest occupied molecular orbital (HOMO) with respect to the Fermi energy of the graphene and the C-AFM metal tip, are determined and discussed with the appropriate electron transport model (double Schottky diode model and Landauer-Buttiker model, respectively) taking into account the voltage-dependent charge doping of graphene. In both types of samples, the energy barrier at the graphene/pentacene interface is slightly larger than that at the pentacene/metal tip interface, resulting in 0.47-0.55 eV and 0.21-0.34 eV, respectively, for the 10-30 nm thick needle-like pentacene islands, and in 0.92-1.44 eV and 0.67-1.05 eV, respectively, for the 2L-3L thick dendritic pentacene nanostructures. We attribute this difference to the molecular organization details of the pentacene/graphene heterostructures, with pentacene molecules lying flat on the graphene in the needle-like pentacene nansotructures, while standing upright in 2L-3L dendritic islands, as observed from Raman spectroscopy.

**Introduction.**

The graphene/semiconductor interface is one of the basic building blocks for hybrid devices and technologies and the detailed knowledge of its electronic transport properties is a key point for developing novel devices. For example, the use of graphene as electrodes in various semiconductor (organic as well as inorganic) devices (transistor, photodetector, molecular-scale junctions, and more...) is expanding.[1-12] In particular, the electronic properties of the graphene/semiconductor device differs from the classical metal/semiconductor Schottky diode. Owing to the zero gap and the low density of states of graphene at the Dirac point, this system has a tunable barrier height, i.e. this barrier depends on the applied voltage. This new behavior (also referenced to as "barristor" for "variable barrier transistor")[13-16] makes the graphene/semiconductor interface a very interesting platform for the study of electronic transport at the interface and for novel devices.[7, 8, 17] However, the results on systems combining graphene with organic semiconductors (OSC) remains scarce to date.[2-4, 8, 14, 17-25] Few studies have shown this Gr/OSC barristor effect for several p-type OSC, e.g. pentacene,[14, 21, 24, 25] benzothienol-benzothiophene (BTBT),[22] dinaphtol-thienol-thiophene (DNTT),[2] poly(3-hexylthiophene) (P3HT),[4, 12] and n-type OSC such as $C_{60}$ and derivatives.[3, 4, 23] Several issues remain however to be more deeply investigated. For instance, all these results were obtained so far for macroscopic diodes with lithographed electrodes (few hundreds of μm) and relatively thick OCS films (>200 nm). Furthermore, an understanding of the role of the molecular organization at the nanoscale is still lacking for these devices.

In this work, we study the charge transport mechanism across a pentacene/graphene (P5/Gr) interface for CVD Gr on $SiO_2$/Si substrates and two different P5 nanostructures: two-three layers (2L-3L) thick dendritic islands and needle-like P5 islands, 10-30 nm thick and ~ μm long. By conductive-AFM, we systematically report for both types of samples almost symmetric current-voltage behaviors against the rectifying (diode) behavior observed on macroscopic Gr/P5 devices.[2-4, 14, 21-25] The energy barrier heights at the interface and the molecular orbital energies in the P5 nanostructures are determined and discussed by fitting the experimental data with appropriate analytical electron transport model, i.e. a double Schottky diode model for the needle-like P5 nanostructures, and the Landauer-Buttiker model for the few layers nanostructures. In these models, we have introduced a suitable analytical formulation of the voltage-dependent charge doping in the graphene. The difference in the interface energetics is likely related to the P5/Gr interface micro- and nano-structuration: Raman spectroscopy indicates that the P5 molecules lie flat on Gr for the 10-30 nm thick P5 nanostructures, while they stand upright in the 2L-thick P5 islands. The P5 molecular orientation depends on the cleanliness of Gr and as such, it can be controlled by annealing the Gr at different temperatures.

**Graphene/pentacene device fabrication and characterization.**

To study the relationship between the molecular structure and the electrical properties of Gr/OSC interface, we fabricated samples on which micro-Raman spectroscopy measurements, topographic tapping-mode atomic force microscopy (TM-AFM) and conductive-AFM (C-AFM) can be done on the same device. The samples provide an easy way to allow the characterization by the three methods at the same position, which will help to correlate the structural and electronic properties. Figure 1 illustrates the process of the sample fabrication. The electrodes are first fabricated (e-beam lithography) on a 285 nm thick $SiO_2$ thermally grown on Si (see section 1 in the Supporting Information). Then, the CVD graphene is transferred via a wet transfer protocol (section 2 in the Supporting Information),[26-30] and cleaned by annealing in $N_2$ for 3h at 150°C or 450°C to remove PMMA residues. Different annealing temperatures are used to achieve different levels of Gr cleanliness and therefore different P5 molecular orientation, as discussed in the following. The Gr flakes are examined by tapping-mode AFM (TM-AFM) and micro Raman spectroscopy (see section 2 in the Supporting Information). Finally, the OSC (pentacene) is deposited (section 3 in the Supporting Information) by evaporation on the clean graphene surface in a vacuum chamber ($\sim 10^{-6}$ mbar). The detailed fabrication protocol is described in the Supporting Information.

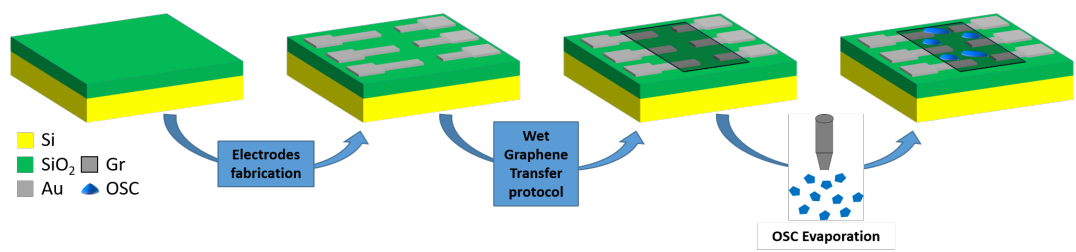

*Figure 1. Scheme of the sample fabrication. The process consists of 3 steps: (i) the fabrication of the electrodes, (ii) the transfer of graphene and (iii) the evaporation of OSC nanostructures onto the graphene.*

By tuning the Gr annealing temperature (see the Supporting Information), we fabricated two types of P5 nanostructures on Gr.[1, 31, 32] Sample A: needle-like (∼ few μm long, ∼ few hundreds nm large and ∼ 10-30 nm height) P5 nanostructures (N-like P5). Sample B: ultra-thin dendritic islands (2L to 5L thick) of P5.

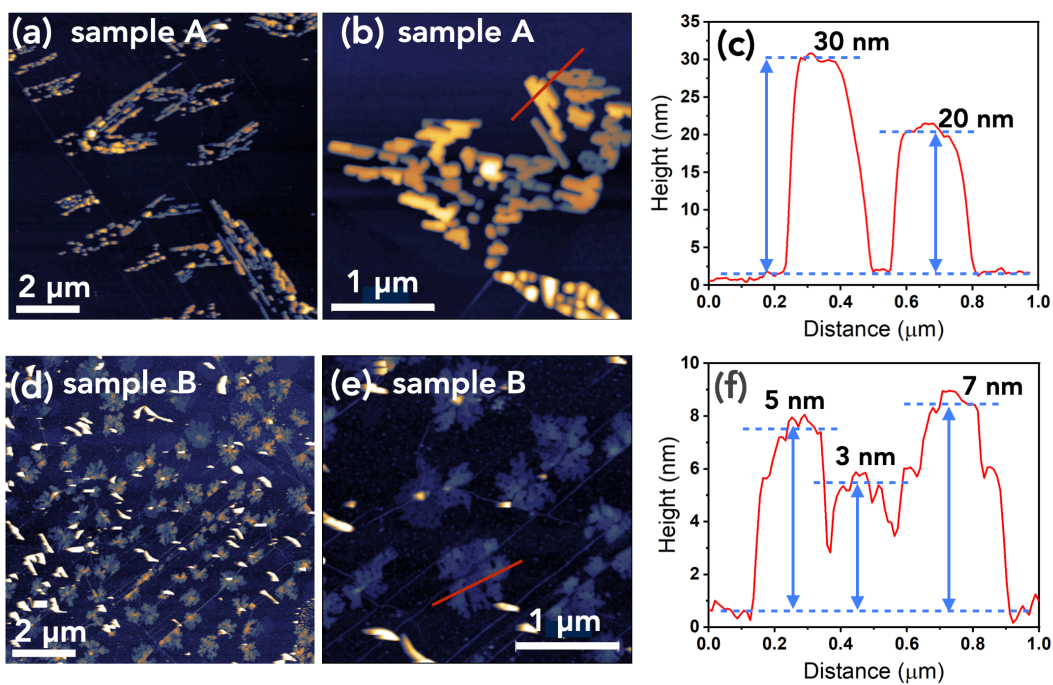

*Figure 2.* TM-AFM images (10 μm x 10 μm) and (3 μm x 3 μm) of sample A (a,b) and sample B (d,e). Height profiles along the red lines for (c) sample A and (f) sample B.

Figure 2 shows the TM-AFM images and typical height profiles for the two samples. These images show the different islands shape grown on the Gr flakes annealed at different temperatures $T_a$. For $T_a=150°C$ (sample B), we mainly obtained dendritic shape of P5 islands with a thickness of ∼ 3-10 nm, which indicate few layers of P5 molecules. Whereas, on Gr annealed at $T_a=450°C$ (sample A), we obtained needle-like shaped islands with a thickness going from 15 to 30 nm indicating several layers of P5, in agreement with literature.[1, 31, 32] The two Gr annealing temperatures were chosen to obtain the desired P5 structures: 2D growth of monolayers on as-transferred Gr (or annealed at low temperature) on $SiO_2$ vs. 3D growth of N-like islands on Gr annealed at >350°C) according to literature.[1, 31, 32] These changes in the morphology of the P5 islands are likely linked to the presence of PMMA residues, as well as defects in the Gr layer and Gr doping.[33-35] Finally, as a reference and for the purpose of comparison with the results reported in the literature for macroscopic Gr/P5 diodes (with large lithographed electrodes and thicker films),[14, 21, 24, 25] we evaporated a 400 nm thick P5 on Gr ($T_a$=150°C) through a shadow mask (lateral size, typically 100 µm x 100 µm), see details in section 4 in the Supporting Information.

| | sample A | sample B |
|---|---|---|
| shape | needle-like | few-layers (2-3) island |
| thickness | 10-30 nm | <10 nm |
| lateral size | ~ few µm long, ~few hundreds nm large | ~ 1 µm (diameter) |
| interface structure | 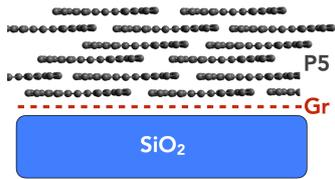 | 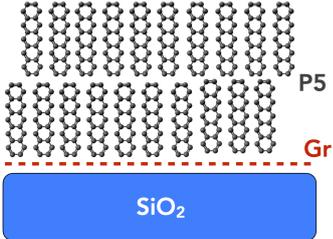 |

*Table 1. Summary of the main characteristics of the two samples: shape, thickness and typical lateral size of the P5 nanostructures deposited on Gr (from AFM measurements, see text) and schematic view of the Gr/P5 interface (from Raman spectroscopy, see text).*

Figures 3a and 3b show the P5/Gr heterostructure and Gr Raman spectra of sample A and sample B, respectively. We identify several Raman peaks assigned to the P5 molecules.[36] The features at 1533, 1457, 1409, 1371, 1178, and 1158 cm$^{-1}$ are assigned to the $A_g$ fundamental band, and the band at 1596 cm$^{-1}$, just before the G band of graphene at 1608 cm$^{-1}$ (Fig. 3c) is assigned to a $B_{3g}$ fundamental band. These Raman features originate from various vibrational modes of the C–H and C–C bonds of P5. The peaks at 1158 and 1178 cm$^{-1}$ are associated with the displacement of H atoms located at the ends and sides of the P5 molecule. For the C–C stretching modes, the peaks are assigned as C–C short axis (1371 and 1533 cm$^{-1}$) and C–C long axis (1596 cm$^{-1}$) modes. The $B_{3g}$ bands (C-C long axis) have low Raman intensities when the long axis of a P5 molecule is perpendicular to the electric field vector of the laser (i.e. perpendicular to the surface).[36-41] Therefore, the ratio of the peak intensity R = $I_{1596}/I_{1533}$ can be used to qualitatively define the orientation of the long axis of the P5 molecules on the substrate. For R>1, the molecules are nearly planar oriented and the long axis is parallel to the surface, while for R ~ 0 the molecules are perpendicular to the surface. For sample A (Fig. 3c), the Gr G-peak and the P5 $B_{3g}$ C-C long axis peak (1596 cm$^{-1}$) are clearly distinguishable, and we found an average of R ~ 1.5 that indicates a preferential planar orientation of the molecules (Table 1). It is worth observing that the graphene G-peak of sample A is blue-shifted, as well as the 2D peak (see section 2 in the Supporting Information). This is typically due to p-doping,[42] which is known to occur for graphene on $SiO_2$ substrates after annealing at high temperatures.[1, 43, 44]

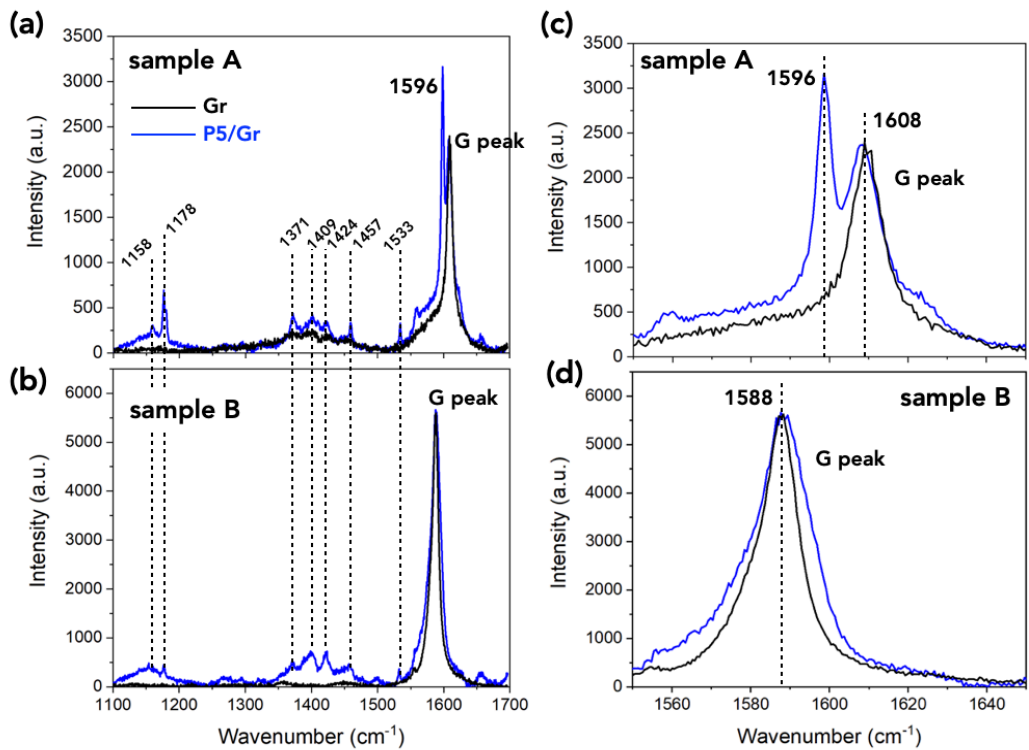

***Figure 3.*** *(a) Raman spectra of P5 islands (blue curve) grown on Gr flake annealed at 450°C (sample A) and of the annealed Gr alone (black curve). (b) Raman spectra of P5 islands (blue curve, sample B) grown on Gr flake annealed at 150°C (black curve). (c) Zoom of the G-peak and $B_{3g}$ P5 peak region (sample A) to highlight the P5 contribution. (d) Same as panel (c) for sample B. The full width at half maximum (FWHM) of the G peak is 11 cm$^{-1}$ (Gr alone) and 19 cm$^{-1}$ (Gr/P5) for sample A and 12 cm$^{-1}$ (Gr alone) and 19 cm$^{-1}$ (Gr/P5) for sample B.*

For Sample B, the Gr G-peak and the P5 $B_{3g}$ C-C long axis peak are not clearly distinguishable (Fig. 3d): either they are one on top of the other, or the P5 $B_{3g}$ C-C long axis peak is absent or very weak and the peak at ca. 1588 cm$^{-1}$ is mostly due to the graphene G-peak. In the former case, we could not conclude anything on the P5 molecular orientation, while in the latter case R ≈ 0 and the P5 molecules are preferentially oriented upright on the Gr surface. The peaks have similar intensities (Fig. 3d). Since the pentacene has an optical absorption coefficient of $\sim 10^5$ cm$^{-1}$ at 473 nm (excitation light, see Methods),[45] the absorption by a 3-5 nm thick P5 is negligible and it is unlikely that the observed peak at 1588 cm$^{-1}$ can be the superposition of the G-peak and B$_{3g}$ P5 peak (or this latter should be extremely weak). Therefore, we assume that the peak at 1588 cm$^{-1}$ is mostly due to the G-peak of graphene (see also Fig. S2 in the Supporting Information) and that the P5 B$_{3g}$ peak is not observed in agreement with what was previously reported in the literature for similar systems, and thus the P5 molecules are preferentially standing upright on the Gr surface (Table 1).[1] For the two samples, the FWHM of the Gr G-peak slightly increases after the deposition of P5 (from 11-12 cm$^{-1}$ to 19 cm$^{-1}$). It is known that the G-peak FWHM depends on the charge density in Gr,[42, 46, 47] and on the local inhomogeneity of the charge density within the laser spot size ($\sim 1$ μm).[47-49] Therefore, the observed increase of the FWHM after the P5 deposition may be due to locally inhomogeneous charge transfer at the P5/Gr interface (in relation with the likely inhomogeneous organization of the P5 molecules deposited on the Gr) leading to an increase of local charge density fluctuations in the Gr.

**Electron transport at the graphene/pentacene interface.**

We measured the current-voltage (I-V) characteristics of the Gr/P5 heterostructures at the nanoscale using C-AFM as depicted in Fig. 4a. Few tens of I-V were acquired on the P5 islands and plotted (gray lines) on Fig. 4b-d for the N-like P5 islands with 3 different thicknesses (from profiles by TM-AFM, see typical examples in Fig. 2). The blue line is the mean $\bar{I}$-V curve. I-V traces at the sensitivity limit of the C-AFM trans-impedance preamplifier ($\sim 10^{-13}$-$10^{-12}$ A) or displaying large noise (e.g. sudden large jumps of current) have been discarded (see C-AFM method, section 5 in the Supporting Information). We note that, in all measurements, V is the applied voltage between the C-AFM tip and the Au electrode and we neglected the voltage drop in the Gr series resistance (between the P5 island and the Au electrode, Fig. 4a) since this series resistance is low (~5 kΩ, see Fig. S5 in the Supporting Information) leading to a small voltage drop (< 50 μV) with a maximum measured current of ~$10^{-8}$ A (Figs. 4 and 5).

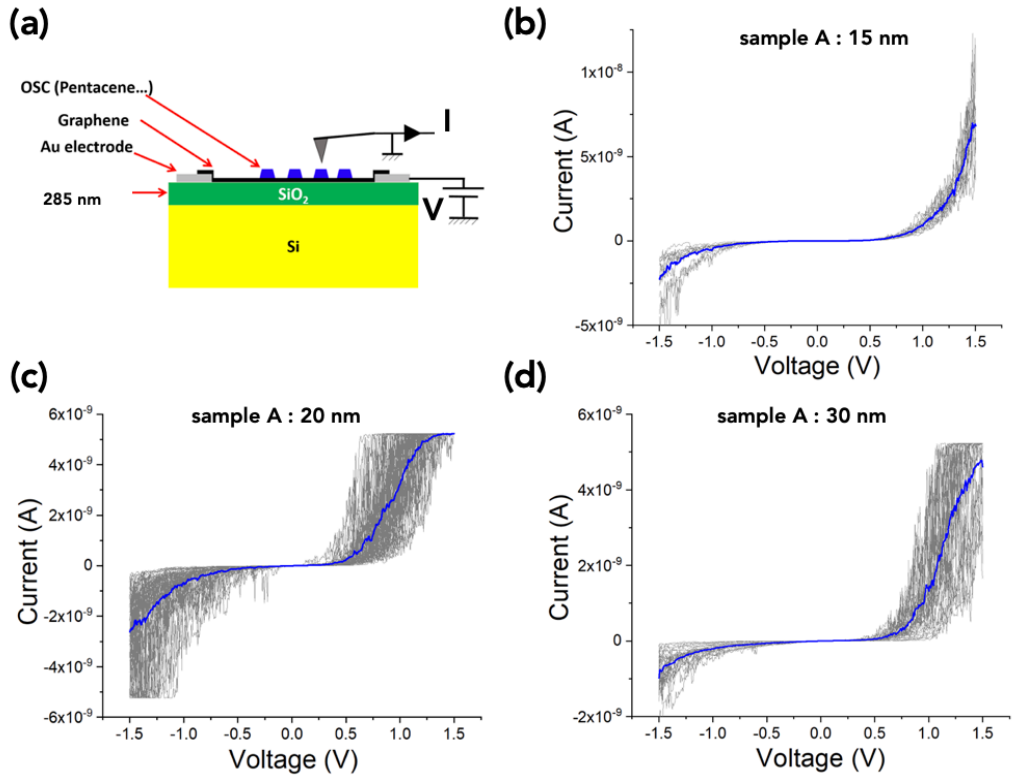

*Figure 4.* (a) Scheme of the Gr/P5 heterostructures on a SiO$_2$/Si substrate with Au electrodes contacting the Gr flakes and the PtIr tip of C-AFM on the P5 island. Plots of the current-voltage curves measured with the C-AFM tip located on different P5 nanostructures, the voltage is applied on one of the lateral Au electrodes connecting the Gr flake: (b) sample A, 15 nm thick N-like P5 island, 13 I-V traces; (c) sample A, 20 nm thick N-like P5 island, 52 I-V traces and (d) sample A, 30 nm thick N-like P5 island, 28 I-V traces. The blue lines are the mean $\bar{I}$-V curves. The plateau at ±5x10$^{-9}$ A at |V|≳1V in several I-V traces corresponds to the compliance limit of the C-AFM trans-impedance preamplifier.

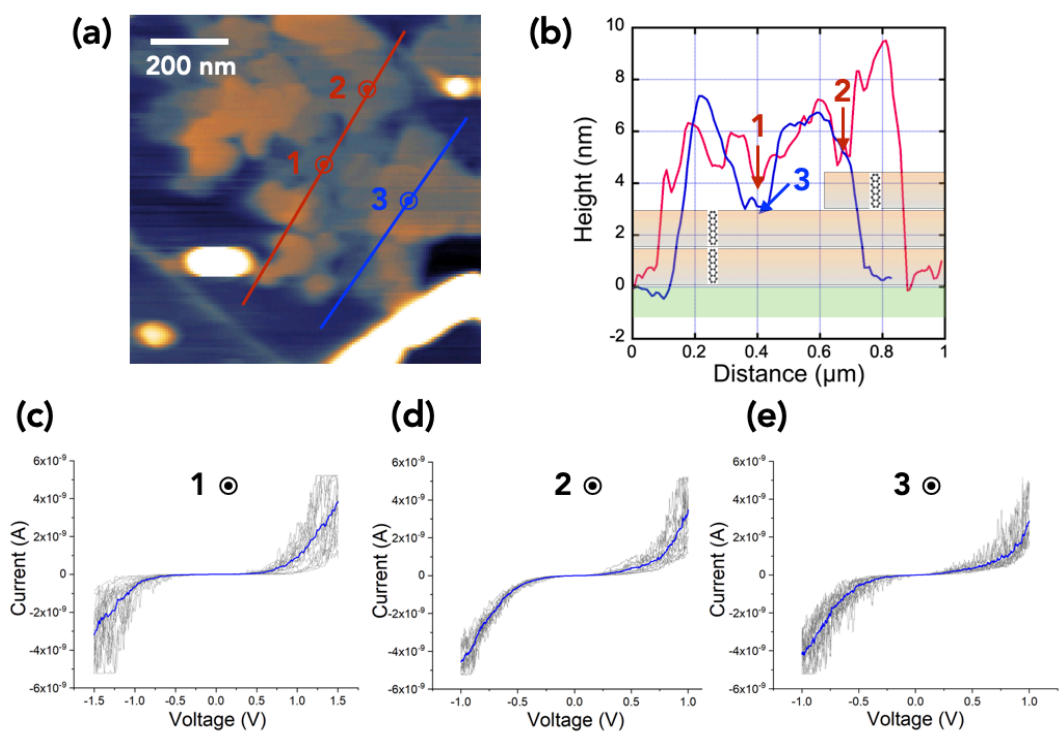

*Figure 5.* (a) Topographic (tapping mode) AFM image of the few-layer P5 islands (sample B) and (b) height profile with the location of points #1, #2 and #3 (circled bullets in (a)) around where the I-V datasets are measured by C-AFM (see Methods). The light green bar corresponds to the baseline (Gr substrate), the light orange/gray bars symbolize the P5 monolayers (1.5 nm thick). (c-e) I-V curves (gray lines) and mean Ī-V curves (blue lines), for the 3 datasets measured at points #1 (20 I-V traces), #2 (20 I-V traces) and #3 (18 I-V traces).

Figure 5 gives the electron transport properties of the few-layer thick P5 dendritic islands (sample B). We have measured the I-Vs at three locations (circled bullets in Fig. 5a, seeMethods and herei sn referred to as points #1, #2 and #3, respectively). From the height profiles (Fig. 5b), these Gr/molecules/PtIr junctions correspond to about 2L of pentacene (points #1 and #3) with a thickness of ≈3 nm (point #3) and ≈4 nm (point #1) (≈1.5 nm for the length of the P5 molecules, standing upright on the Gr surface as determined by Raman spectroscopy ( *vide supra,* Fig. 3), while we have 3L for the measurements done at the point #2.

For the N-like P5 islands (sample A) and few-layer P5 islands (sample B), almost weak asymmetric I-V traces are systematically measured, while the I-V curves for a 400 nm thick P5 film (used as a reference sample, Fig.S3 section 4 in the Supporting Information) clearly show the expected diode behavior (forward regime at V<0, and reverse regime at V>0 on the Gr electrode, PtIr tip on P5 grounded) with a diode rectification ratio, R = $\bar{I}(V=-1.5V)/\bar{I}(V=1.5V)$, is ≈ $10^2$ in agreement with reported results on macroscopic P5/Gr diodes (see more details in the Supporting Information, section 4).[14, 21, 24, 25]

Thus, we infer that different electron transport mechanisms are involved in the nanoscale Gr/P5/metal devices reported above, which we discussed below.

**Discussion.**

Since the transport physics depends on the device scale, we analyzed the I-V datasets with different transport models as the function of the type of P5 nanostructures.

*Few layers (2L-3L) pentacene/graphene devices.*

We first discuss the few layers (2L-3L) P5 devices (sample B, Fig. 5), which can be considered as molecular junctions. We used a molecular-based model (2-site model), considering two molecular orbitals (HOMO of the pentacene) in series between the Gr and PtIr tip (Fig. 6b for the equilibrium situation at V=0). The model is parametrized by the energy levels $\varepsilon_1$ and $\varepsilon_2$ (with respect to the Fermi energy), by the electronic coupling energy Γ between the P5 molecules and the electrodes and τ the intermolecular coupling between the two molecules. The use of the same coupling term, Γ, while we have two different electrodes is discussed hereafter. In this model, the current-voltage behavior is given by:[50, 51]

$$I(V) = N\frac{e}{h}\frac{\Gamma(2\tau)^2}{\Delta^2 + \Gamma^2}\left[\arctan\left(\frac{\frac{1}{2}eV_i - \varepsilon_{1*}}{\Gamma/2}\right) + \arctan\left(\frac{\frac{1}{2}eV_i + \varepsilon_{1*}}{\Gamma/2}\right) + \frac{\Gamma}{2\Delta}ln\left(\frac{\left(\frac{1}{2}eV_i - \varepsilon_{1*}\right)^2 + (\Gamma/2)^2}{\left(\frac{1}{2}eV_i + \varepsilon_{1*}\right)^2 + (\Gamma/2)^2}\right)\right.$$

$$\left. + \arctan\left(\frac{\frac{1}{2}eV_i - \varepsilon_{2*}}{\Gamma/2}\right) + \arctan\left(\frac{\frac{1}{2}eV_i + \varepsilon_{2*}}{\Gamma/2}\right) - \frac{\Gamma}{2\Delta}ln\left(\frac{\left(\frac{1}{2}eV_i - \varepsilon_{2*}\right)^2 + (\Gamma/2)^2}{\left(\frac{1}{2}eV_i + \varepsilon_{2*}\right)^2 + (\Gamma/2)^2}\right)\right]$$

(1)

with e the electron charge, h the Planck constant and N≈60 the number of molecules contacted by the C-AFM tip (see section 5 in the Supporting Information). The parameters $\varepsilon_{1*}$ and $\varepsilon_{2*}$ are two voltage-dependent effective energy positions of the molecular orbitals as described in the Supporting Information (section 7). In Eq. (1), $V_i$ is the internal voltage between the Dirac point of the Gr electrode and the Fermi energy of the PtIr tip which is related to the applied voltage V by (Figs. 6a and c)

$V = V_i + \delta(V)/e$  (2)

where $\delta(V)$ is the energy separation between the Dirac point and the Fermi energy of the Gr electrode due to voltage-induced doping of the Gr layer, $\delta = \varepsilon_{DP} - \varepsilon_F$. This quantity is calculated versus the applied voltage V according to Feenstra et al.[52] and $\delta(V)$ varies between -0.2 and 0.2V in the applied voltage range (Fig. S6, section 7 in the Supporting Information for details).

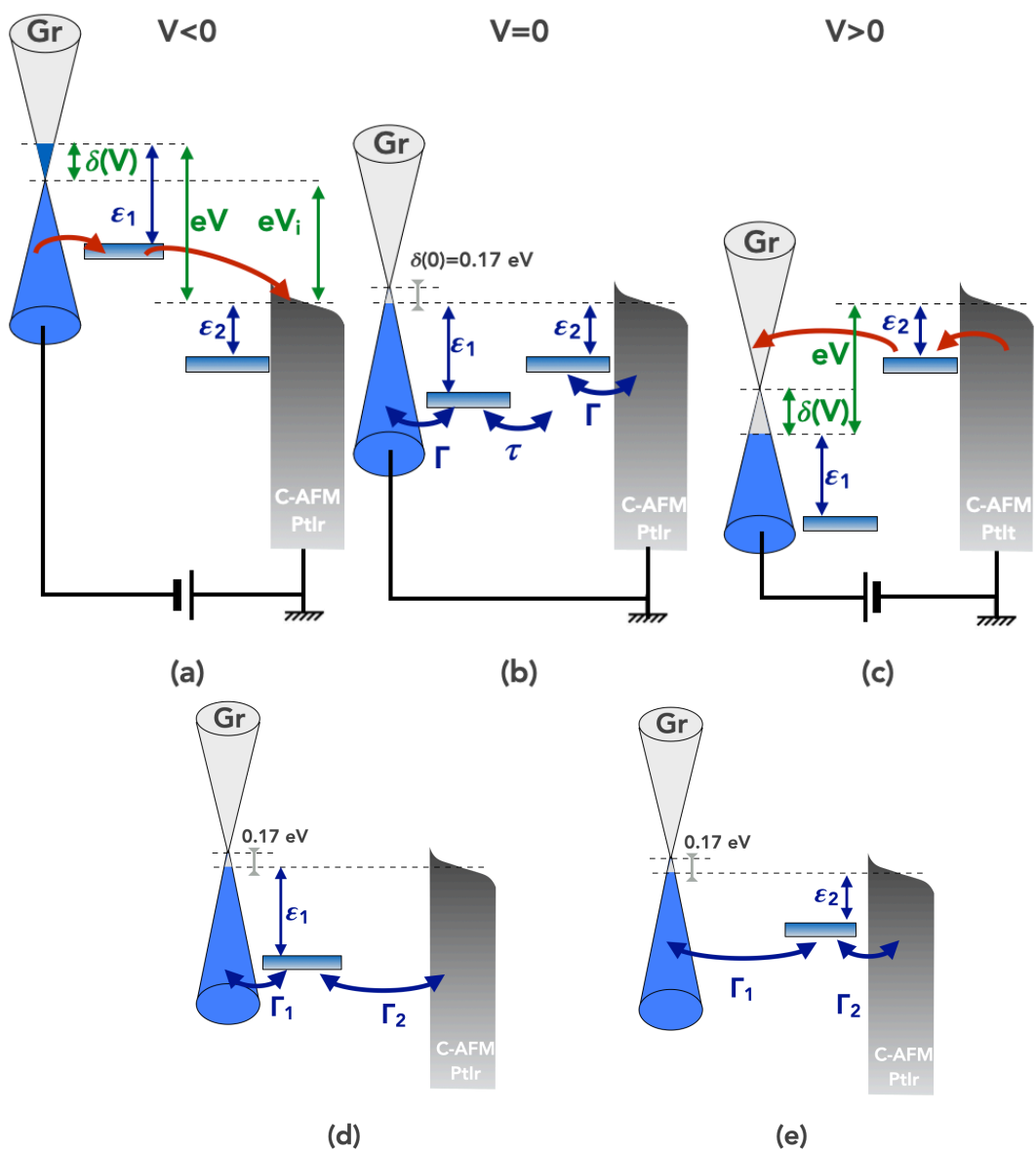

*Figure 6. Proposed energy diagram of the Gr/bilayer P5/PtIr tip junction at (a) negative applied voltage V, (b) at equilibrium (V=0) and (c) at positive applied voltage V. $\varepsilon_1$ and $\varepsilon_2$ are the energy positions of the HOMO of P5 in each monolayer (with respect of the Fermi energy). The Gr layer is slightly hole doped at V=0, $\varepsilon_{DP} - \varepsilon_F$ = 170 meV, see section 2 in the Supporting Information), $\Gamma$ is the electronic coupling energy between the P5 molecules and the electrodes and $\tau$ is the intermolecular coupling between the two molecules. With an applied voltage*

*V (panels a and c), the Fermi energy of Gr is shifted by δ(V), $V_i$ is the internal voltage between the Gr Dirac point and the Fermi energy of the PtIt tip. Equivalent energy diagram of the single energy level model (SEL) used to fit the I-V curve by part at V<0 (panel d) and at V>0 (panel e).*

Figure 7a shows the fit of this model on the mean Ī-V measured at point #3 with the following fitted parameters : $\varepsilon_1$ = 0.92 eV, $\varepsilon_2$ = 0.82 eV, Γ = 0.19 eV and τ = 7 meV. We have also fitted all the individuals I-Vs of the dataset shown in Fig. 5 and extracted the statistical distributions of the 4 parameters, Figs 7b and c for data at point #3. The same analysis on the mean Ī-V at points #1 and #2 are given Figs. S8 and S9 in the Supporting Information and the parameters are summarized in Tables 2 and S1.

In the three cases, the trends are similar: a slightly larger value for $\varepsilon_1$ than for $\varepsilon_2$, a strong electronic coupling Γ to the electrodes and a weak intermolecular coupling τ between the adjacent P5 monolayers (the case of the results at point #2 with 3L of P5 instead of 2 is discussed below). The fact that the two monolayers of P5 are energetically weakly coupled, while the molecules are strongly coupled to the electrodes, allows us to consider that the molecular junction can be viewed as two blocks, a Gr-P5 molecule on one side and a P5-PtIr on the other side, and that under the applied voltage these two blocks are energetically shifted with respect to each other (Figs. 6a and c). Consequently, at V>0, the Gr-$\varepsilon_1$ block is shifted downward and only the HOMO level $\varepsilon_2$ contributes to the electron transport (Fig. 6c), while this is the HOMO level $\varepsilon_1$ at V<0 (Fig. 6a). To check this interpretation, we fitted separately the positive and negative voltage parts of the mean Ī-V with a single energy level (SEL) model considering that only one molecular orbital contributes to the electron transport, $\varepsilon_1$ at V<0 as depicted in Fig. 6d and $\varepsilon_2$ at V>0, Fig. 6e (see details in Section 7 in the Supporting Information). Figure 7d shows that this modified SEL model perfectly fit the data with $\varepsilon_1$ = 0.95 eV (at V<0) and $\varepsilon_2$ = 0.78 eV (at V>0). These results support the energy level values directly given by the two-site model. The same behaviors and conclusions are obtained for the measurements at points #1 and #2 (see Figs. S8 and S9, section 8 in the Supporting Information) as summarized in Tables 2 and S1. We note that the fit of the 2-site model on the mean $\bar{I}$-V measured at point #2 is a bit worse than at points #1 and #3 (Figs. S8 and S9, Supporting Information), nevertheless, the same trends are observed as for the 2L molecular junctions (Supporting Information, section 8). We also note that the coupling energies $\Gamma_1$ and $\Gamma_2$ are almost similar for the two electrodes (Table S1 in the Supporting Information) and also of the order of magnitude as the parameter $\Gamma$ in Eq. (1), justifying this simplification of using a single parameter $\Gamma$ in the 2-site model.

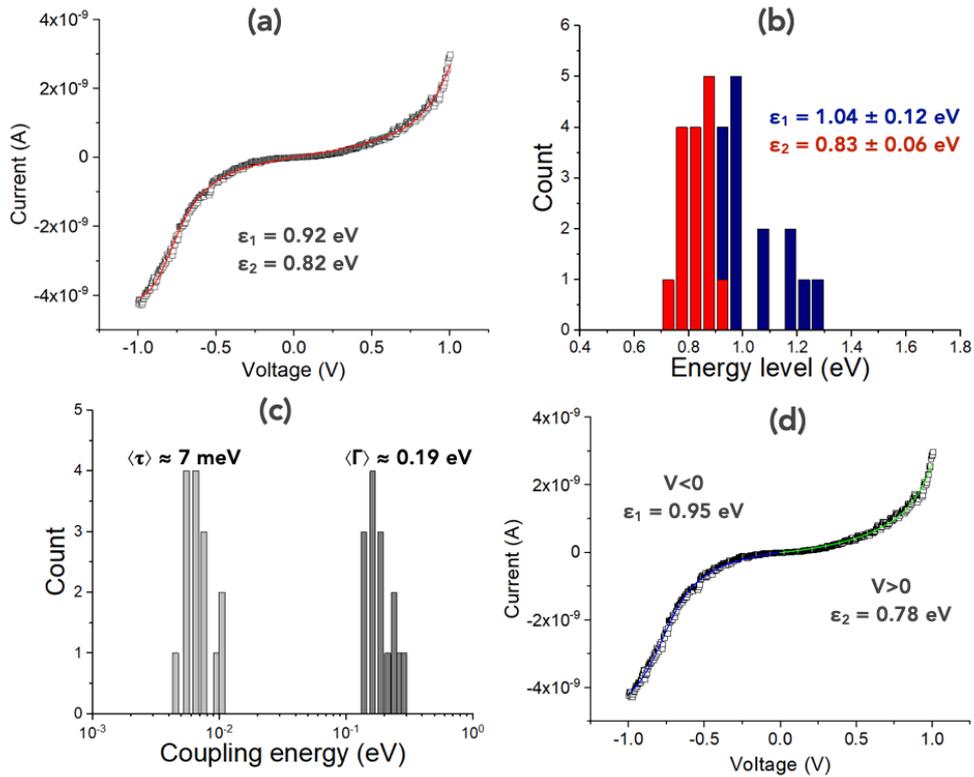

*Figure 7. Mean $\bar{I}$-V of the dataset measured at point #3 (from Fig. 5), open black squares: (a) fit (red curve) with the 2-site model taking into account the voltage-induced charging of the Gr layer (Eqs. 1-2); (b) statistical distribution of the*

*energy levels ε₁ and ε₂ (arithmetic mean ± standard deviation); (c) statistical distribution of the coupling energies Γ, τ (arithmetic mean); (d) fit of the mean Ī-V with the modified SEL (Eq. S7 in the Supporting Information) for the positive voltage (green curve) and negative voltage (blue curve). The fitted energy levels of the molecular orbitals (energy scheme in Fig. 6), ε₁ and ε₂, are given on the figures. The other fit parameters (Γ, τ, Γ₁ and Γ₂) are given in Table S1 (Supporting Information). The fits for the I-Vs measured at points #1 and #2 are given in the Supporting Information.*

|    |            | 2-site model | SEL (V < 0) (fit on mean Ī-V) | SEL (V>0) | 2-site model (statistics) |
|----|------------|--------------|-------------------------------|-----------|---------------------------|
| #1 | $\varepsilon_1$ (eV) | 1.32 | 0.96 | n.a. | 1.28 ± 0.16 |
|    | $\varepsilon_2$ (eV) | 0.73 | n.a. | 0.63 | 0.78 ± 0.11 |
| #2 | $\varepsilon_1$ (eV) | 1.17 | 0.99 | n.a. | 1.12 ± 0.18 |
|    | $\varepsilon_2$ (eV) | 0.89 | n.a. | 0.73 | 0.94 ± 0.11 |
| #3 | $\varepsilon_1$ (eV) | 0.92 | 0.95 | n.a. | 1.04 ± 0.12 |
|    | $\varepsilon_2$ (eV) | 0.82 | n.a. | 0.78 | 0.83 ± 0.06 |

***Table 2**. Values of the fitted energy levels (see diagrams in Fig. 6) for the 2-site and SEL models (n.a. stands for non-applicable) on the mean Ī-V. Energy level values from the statistical analysis (arithmetic mean ± standard deviation) for the 2-site model only. From the statistical analysis, we deduce : $0.92 < \varepsilon_1 < 1.44$ eV and $0.67 < \varepsilon_2 < 1.05$ eV.*

**Needle-like pentacene/graphene devices.**
For sample A (10-30 nm thick N-like P5 nanostructures), with more than 2-3 monolayers of P5, a molecular junction model is no longer suitable. Strictly speaking, molecular-scale electron-transport models are applicable for single-molecule or few-molecule junctions (e.g. 10-100 typically),[53] and they are not suitable for devices with a larger number of molecules. If this model is nevertheless used to fit the data of sample A, it gave unrealistic and very large ε values (up to 3 eV, larger than the P5 HOMO-LUMO gap of 2.2 eV, Fig. S7 in the Supporting Information, section 7). We used a double Schottky barrier (SB) model recently developed and demonstrated for two-terminal nanodevices.[54] The model considers two back-to-back Schottky diodes at each side of a resistance (Fig. 8a): one accounts for the Gr/P5 interface and the other one for the interface at the P5/PtIr C-AFM tip, the resistance being the charge transport through the P5 nanostructure. This analytical model has been proven able extracting both the SB energy simultaneously from the fit of I-V curve even when the two SBs are dissimilar (i.e. different energy barriers, ideality factors and contact areas).[54] In this model, the current is always limited by the saturation current of the reverse-biased diode, the other diode being in the forward regime. Here, with a p-type OSC, the voltage applied on the Gr and the PtIr C-AFM grounded, we measure the Gr/P5 diode at V>0 and the P5/PtIr tip diode at V<0 (insets in Fig. 8b). The current is given for a p-type semiconductor by[54]

$$I(V) = \frac{2 I_{S1} I_{S2} \sinh\left(\frac{qV_i}{2kT}\right)}{I_{S1} \exp\left(-\frac{qV_i}{n_1 2kT}\right) + I_{S2} \exp\left(\frac{qV_i}{n_2 2kT}\right)} \quad (3)$$

In this equation, $I_{S1,2}$ are the saturation currents where the subscripts "1" and "2" refer to the Gr/P5 and P5/PtIr tip interfaces, respectively:

$$I_{S1,2} = S_{1,2} A^* T^2 \exp\left(-\frac{\phi_{B1,2}}{kT}\right) \quad (4)$$

with $S_{1,2}$ the contact areas ($S_1 \approx 0.1$ μm$^2$ the surface between Gr and N-like P5 islands, see AFM images, Fig. 2 and $S_2 \approx 15$ nm$^2$, the C-AFM tip contact area, see section S5 in the Supporting Information), k the Boltzmann constant, T=295K and A*=10$^2$ Acm$^{-2}$K$^{-2}$ (the effective Richardson constant for OSC).[55] In Eq. (3), $n_{1,2}$ are the ideality factors and $V_i$ the internal voltage between the Gr Dirac point and the PtIr tip Fermi level as above (Eq. 2). Fig. 8b shows a typical fit of this model on the mean Ī-V of the 15 nm-thick N-like P5 island, from which we clearly extract the parameters of the two dissimilar SBs. The energy barrier height at the Gr/P5, $\Phi_{B1}$, is slightly higher than at the P5/PtIr tip, $\Phi_{B2}$, a result confirmed by the statistical analysis of the I-V dataset (Fig. 8c). We also note a slightly higher ideality factor at the Gr/P5 interface (Fig. 8d). The same trends are observed for the 20 and 30-nm thick N-like nanostructures (Fig. S10 in the Supporting Information). Table 3 summarizes the fitted SB parameters. Globally, to summarize, we get 0.47 < $\Phi_{B1}$ < 0.55 eV ($n_1 \approx$ 1.3-1.74) and 0.21 < $\Phi_{B2}$ < 0.34 eV ($n_2 \approx$ 1.11-1.25). This behavior ($\Phi_{B1}$ > $\Phi_{B2}$) is reminiscent of the observation that $\varepsilon_1$ > $\varepsilon_2$ for the 2L-3L P5 islands and this feature can be accounted for by the slightly larger work function of PtIr C-AFM tip, ≈5 eV,[56] than of Gr (≈ 4.6 eV).[8] The result $n_1 > n_2$ indicates a less ideal Gr/P5 interface than the P5/C-AFM tip one, and it can be related to more defects, impurities at the Gr/P5 interface. Albeit this parameter was not systematically reported in previous works on Gr/P5 Schottky diodes,[14, 21, 24, 25] we note that the present value is better than ≈3 reported for HOPG/P5 diodes.[21]

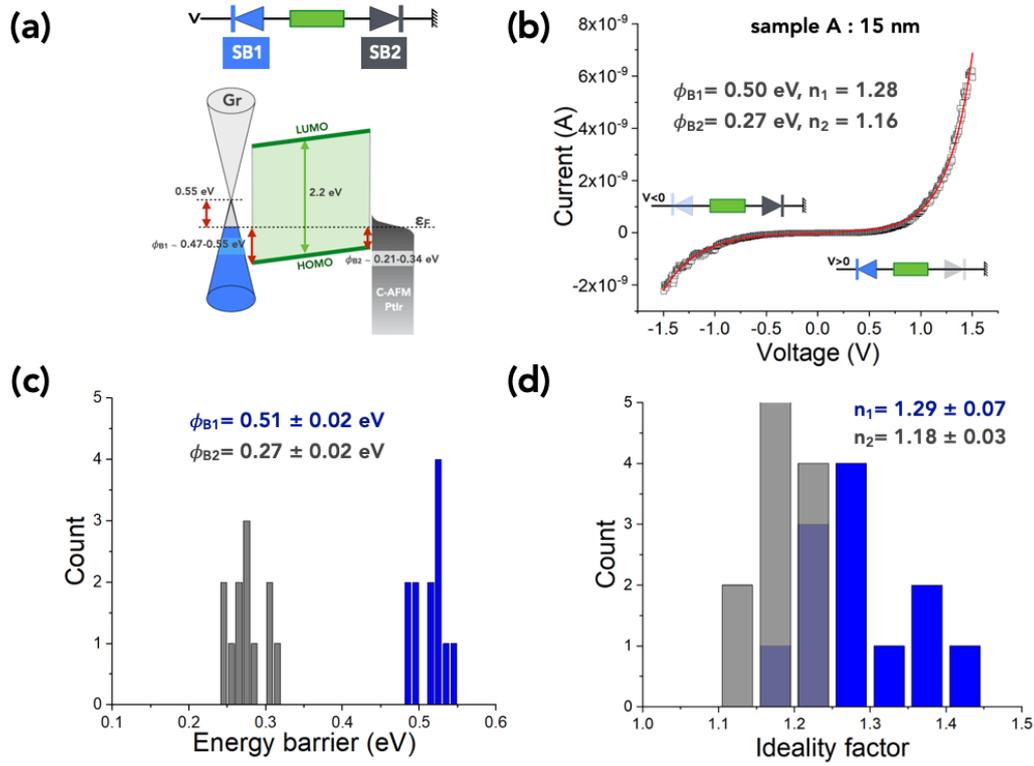

*Figure 8*. Fits of the double Schottky diode model on the I-V dataset of sample A. (a) Scheme of the electronic structure of the Gr/P5/PtIr structure and the two Schottky diodes in opposite direction at zero bias; the Gr layer is hole doped, $n_0$ = $1.8 \times 10^{13}$ cm$^{-2}$, at V=0, $\varepsilon_{DP}$ at 0.55 eV above the Fermi energy, see section 2 in the Supporting Information); (b) fit (red line) on the mean $\bar{I}$-V (open square, data from Fig. 4) of the 15 nm thick N-like P5 island. In the inset, the shaded diodes are in the forward regime. At V>0 (V<0, respectively), the measured current is the reverse current of the Gr/P5 (P5/PtIr tip, respectively) diodes; (c) statistical distribution of the barrier heights by fitting the model on all the individual I-V traces of the dataset shown in Fig. 4; (d) statistical distribution of the ideality factors. The fits on the datasets of the 20 and 30 nm thick P5 islands are shown in Fig. S10 in the Supporting Information. All the fit parameters are summarized in

*Table 3. For the fits, in Eq. S6, the capacitance C is 2.9x10$^{-7}$ F/cm$^2$ for the 15 nm thick P5 nanostructures.*

| P5 thickness (nm) | 15 | 20 | 30 |
|---|---|---|---|
| $\Phi_{B1}$ (eV) - fit mean Ī-V<br>$\Phi_{B1}$ (eV) - statistics | 0.50<br>0.51 ± 0.02 | 0.51<br>0.51 ± 0.04 | 0.50<br>0.52 ± 0.02 |
| $\Phi_{B2}$ (eV) - fit mean Ī-V<br>$\Phi_{B2}$ (eV) - statistics | 0.27<br>0.27 ± 0.02 | 0.30<br>0.26 ± 0.05 | 0.29<br>0.31 ± 0.03 |
| $n_1$ | 1.29 ± 0.07 | 1.41 ± 0.28 | 1.50 ± 0.24 |
| $n_2$ | 1.18 ± 0.03 | 1.18 ± 0.07 | 1.15 ± 0.04 |

**Table 3**. *Fitted values of the SB heights at the Gr/P5 ($\Phi_{B1}$) and P5/PtIr ($\Phi_{B2}$) interfaces: values from fits on the mean Ī-V and from statistical data analysis (Figs. 8 and S10) and values of the ideality factors ($n_1$ and $n_2$ from statistical analysis of the dataset) at the same interfaces, respectively.*

More interesting, we systematically obtain higher energy barriers for sample B than sample A : $\varepsilon_1$ > $\Phi_{B1}$ and $\varepsilon_2$ > $\Phi_{B2}$ (see Tables 2 and 3). This result is understood considering a better π-π interaction in sample A between the Gr and P5 molecules with these molecules lying flat on the Gr layer. It is known that stronger π-π interactions in molecular devices lead to i) a reduction of the HOMO-LUMO gap, ii) a broadening of the molecular levels and then a reduced energy difference between the MOs and the Fermi energy.[57] This trend has been observed by ultraviolet photoemission spectroscopy (UPS) with a reduction of the hole injection barrier (HOMO energy relative to Gr Fermi energy) for P5 molecules lying flat on the Gr surface.[1] For the few-layer P5 devices, the HOMO at 0.92-1.44 eV below the Fermi energy of Gr is in agreement with a previous study where values between 0.85 and 1.35 eV were reported depending on how the Gr/P5 heterostructure is coupled to or suspended over (Gr wrinkle) the underlying metal substrate.[58]

**Conclusion.**

In conclusion, the main results of this work are:

- For needle-like pentacene nanostructures on graphene (thickness between 10 and 30 nm and ~few μm long), the zero-bias energy barrier height at the graphene/pentacene interface is 0.47-0.55 eV, slightly larger than at the pentacene/C-AFM PtIr tip interface (0.21-0.34 eV).

- For molecular junctions graphene/2L-3L thick pentacene/C-AFM tip, the HOMO of pentacene at the Gr side is at a deeper energy location (0.92-1.44 eV), while it is at 0.67-1.05 eV for the pentacene contacted by the C-AFM tip.

These differences in the interface energetics are likely related to the details of the pentacene/graphene interface. In the first case, Raman spectroscopy shows that the P5 molecules at the Gr interface are lying flat on the Gr, while they are standing upright in the 2L and 3L thick pentacene islands.

**Methods.**

*TM-AFM and C-AFM.* We measured the topography (tapping mode) and the electron transport properties (C-AFM) with a Bruker ICON machine at room temperature in air (air-conditioned laboratory with controlled $T_{amb}$=22°C and relative humidity of 35-40%). For the TM-AFM, we used a Si cantilevers with a free oscillating frequency $f_0 \approx 370$ kHz and a spring constant k ≈ 42 N/m. AFM images were treated with Gwyddion.[59] For the C-AFM, we used a tip probe in platinum/iridium (with loading force of ≈ 8 nN). The current-voltage (I-V) curves were acquired applying the voltage on the Gr electrode, the C-AFM tip being grounded. We used the C-AFM mode to scan the surface and locate the islands then stabilize the tip on a specific island (known height). On the chosen island, typically 15 to 60 I-V traces were recorded on several points on the same island by moving the C-AFM tip by 50-100 nm in the x-y plane around the same position on the nanostructure. This process was repeated several times on different islands with different thicknesses. Since the I-V measurements were done in air and that oxygen is known to hole-dope the pentacene, the atmospheric effects could be an issue. However, it was shown[21] that, after an initial period of 24h during which weak drift of the I-Vs were observed, the Gr/P5 diodes are stable in air for at least a week (typically the time required to a full C-AFM characterization of one sample). Thus, all the samples were measured under stable environmental conditions and their data can be safely compared.

*Micro-Raman spectroscopy*. We used a LabRAM HR confocal system from Horiba Jobin-Yvon with a 473 nm laser source, a 1800 grooves/mm grating, a spot size of ~1 µm and a resolution of 1 $cm^{-1}$. On each sample, 3 measurements were performed at different locations on the sample and then averaged. All the measurements were done in air at room temperature. Raman data were treated with Labspec5 software provided by Bruker.

## Associated content

The Supporting Information is available free of charge at xxxxxx.

Details on the sample fabrication: CVD graphene growth and transfer, pentacene evaporation; topographic AFM and Raman characterization; reference sample (thick pentacene film); C-AFM protocol; details on the models; additional fits on all the datasets acquired on the samples.

## Author Contributions

M.D.M., J.O., M.S. and D.B. fabricated the samples. R.F. prepared the CVD graphene. M.D.M. performed the Raman spectroscopy, TM-AFM and C-AFM measurements. M.D.M. and D.V. analyzed the data. D.V. and M.C. conceived, supervised the project and secured the funding. The manuscript was written by M.D.M. and D.V. with the contributions of all the authors. All authors have given approval of the final version of the manuscript.

## Note

The authors declare no competing financial interest.

## Acknowledgements.

We acknowledge support of the ANR (France) and SNF (Switzerland), project VdW-OPBT, under grants # ANR-18-CE93-0005-01 and #182544, respectively. We acknowledge D. Vignaud (IEMN) for his help to use the micro Raman spectrometer and discussions on the interpretation of the Raman spectra. The authors also thank the cleanroom operations team of the Binnig and Rohrer Nanotechnology Center (BRNC, IBM-Zurich) and the team of the scanning probe microscopy platform at IEMN for their help and support.

## References.

# Nanoscale electronic transport at graphene/pentacene van der Waals interface.

Michel Daher Mansour[1], Jacopo Oswald[2,3], Davide Beretta[2], Michael Stiefel[2,§], Roman Furrer[2], Michel Calame[2,3,4], Dominique Vuillaume[1].

1. Institut for Electronic, Microelectronic and Nanotechnology, Centre National de la Recherche Scientifique, Villeneuve d'Ascq, France.
2. Transport at Nanoscale Interfaces Laboratory, Swiss Federal Laboratories for Material Science and Technology (EMPA), Dübendorf , Switzerland.
3. Swiss Nanoscience Institute, University of Basel, Switzerland.
4. Department of Physics, University of Basel, Switzerland.

§ Current address : IBM, Binnig and Rohrer Nanotechnology Center, Zurich, Switzerland.
* Corresponding authors: dominique.vuillaume@iemn.fr; michel.calame@empa.ch

## Supporting Information

### 1. Metal electrodes on Si/SiO$_2$.

Ti (5nm) / Au (50 nm) electrodes were fabricated on a 4-inch Si (525 µm thick) / SiO$_2$ (285 nm thick) wafer, which was pre-cleaned in oxygen plasma (600 W for 5 min). The electrodes (Ti/Au) were deposited by e-beam physical vapor deposition (EBPVD) and patterned by liftoff in DMSO (dimethyl sulfoxide) at 100°C for 30 min. The resist for the liftoff (AZ2020nlof) was spin-coated (4000 rpm for 60 s), exposed to UV light (lamp intensity 11 mW/cm$^2$) through an optical mask, and then developed (AZ726mif, 35 s).

## 2. CVD graphene, transfer protocol and characterizations.

Graphene was grown in-house by chemical vapor deposition (CVD) on copper foils. The growth protocol can be found in previously reported works.[1-3] CVD graphene foil (Cu/Gr/PMMA) was placed to float in a copper etchant (Transene CE-100) for 1h, the PMMA layer facing upwards. Once the copper is completely etched (Gr/PMMA), the etchant was removed and replaced with deionized water, twice. Then, the foil was transferred to a 10% HCl cleaning solution for 5 min and transferred back to deionized water, twice. The floating graphene foil (Gr/PMMA) was transferred onto the substrate (Si/SiO$_2$/Ti/Au/Gr/PMMA) and let dry in air for 1 h. The chip was annealed overnight at 80°C in the vacuum (∼1 mbar). Finally, the PMMA is removed in acetone (45 min at 50°C). Fig. S1 shows the TM-AFM of the deposited Gr flakes and after annealing at 150°C and 450°C (3h under N$_2$).

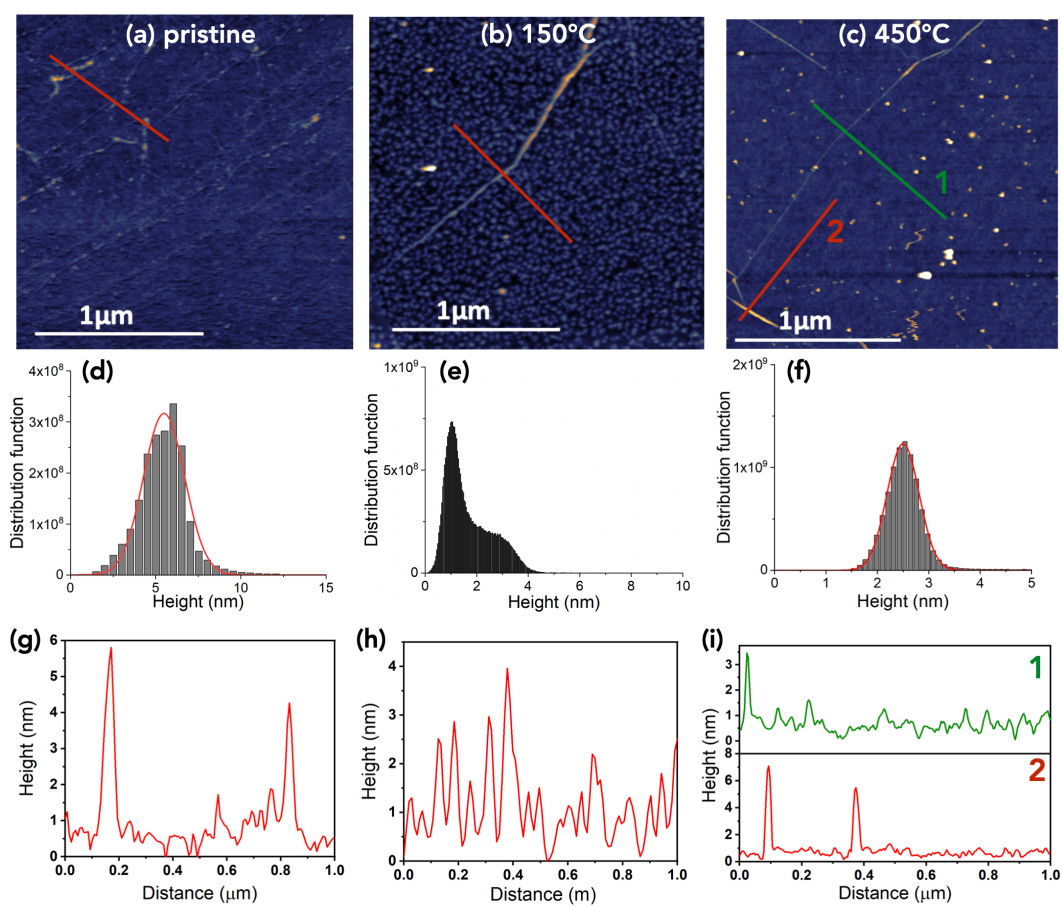

**Figure S1**. TM-AFM images of Gr flakes: (a) after transfer onto SiO$_2$, (b) annealed at 150°C and (c) annealed at 450°C. (d-f) Corresponding histograms of heights (the red lines are Gaussian fits) and (g-i) typical profiles along the lines shown in panels (a-c).

These measurements show a clear removal of PMMA residues after the annealing at 450°C as indicated by a significant decrease of the width of the height distribution ( 2.9 nm FWHM for the as-deposited Gr and 0.7 nm for the Gr flake annealed at 450°C). The PMMA removal is incomplete after the annealing at 150°C (FWHM of 0.74 nm for the main peak, but a larger shoulder is still visible). Wrinkles are still present after the annealing. The calculated rms roughness (masking the wrinkles and the brighter spots probably corresponding to dust since the AFM measurements were done in air) decrease from 2.7 nm for the as-transferred Gr to 1.8 nm after a 150°C annealing and 0.7 nm after a 450°C annealing. The reduction of roughness and height distribution FWHM indicates the removal of PMMA residues from the Gr surface (only incomplete for the annealing at 150°C) and thus a cleaner and smoother surface, which is mandatory to grow the OSC molecules.

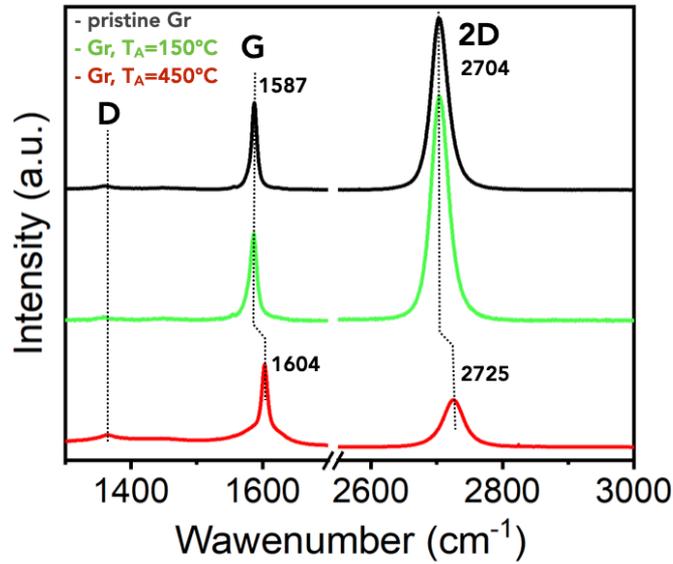

*Figure S2*. *Raman spectra (D, G and 2D peaks) of the Gr flakes after deposition and annealing at 150 and 450°C.*

Fig. S2 shows the Raman spectra of graphene. The annealing at 150°C does not induce significant change. The values of the G and 2D peaks (at 1587 cm$^{-1}$ and 2704 cm$^{-1}$, respectively) slightly higher than the expected values for a neutral Gr indicates a weak residual hole doping.[4, 5] We can estimate the Gr charge density, n, from the relations:[4, 6]

$$\omega_G - \omega_G^0 = (42\ cm^{-1}/eV)|\varepsilon_F|\ ;\ \varepsilon_{DP} - \varepsilon_F = sgn(n)\hbar v_F\sqrt{\pi |n|} \tag{S1}$$

with $\omega_G$ the G peak position, $\omega_G^0$ = 1580 cm$^{-1}$ for the undoped Gr,[4] $\varepsilon_F$ the Fermi energy and $\varepsilon_{DP}$ the energy of the Dirac point, $\hbar$ the reduced Planck constant and $v_F$ the graphene Fermi velocity ~10$^6$ m/s. With $\omega_G$ = 1587 cm$^{-1}$, we get $\varepsilon_{DP}$ - $\varepsilon_F$ = 170 meV and a hole density n = 1.6x10$^{12}$ cm$^{-2}$.

The treatment at 450°C leads to a blue shift of the graphene peaks. The, G and 2D peaks (at 1587 cm$^{-1}$ and 2704 cm$^{-1}$, respectively for the as-transferred Gr and annealed at 150°C) shifted toward higher wavenumbers by ~21 cm$^{-1}$ and ~17 cm$^{-1}$, respectively, for the Gr flakes annealed at 450°C. The intensity ratio 2D/G also decreases from ~2-2.5 (pristine Gr and annealed at 150°C) to ~0.6 after annealing at 450°C, indicating an additional hole doping of the 450°C annealed Gr,[5, 7, 8] likely due to the thermal activation of electron transfer from Gr to the substrate.[9] With Eq. S1, we get $\varepsilon_{DP}$ - $\varepsilon_F$ = 550 meV and a hole density n = 1.8x10$^{13}$ cm$^{-2}$. In all cases, the intensity of the D peak remains low (no defect induced in the Gr).

3. **Pentacene evaporation.**

The purified pentacene (99.999 %), purchased from Tokyo Chemical Industry (TCI), is sublimated by Joule heating (crucible temperatures 90-120°C) in a 10$^{-6}$ mbar vacuum evaporation system (Edwards Auto306) placed inside a glovebox (MBRAUN, H$_2$O and O$_2$ levels below 5 ppm). The deposition rate and time (typically ≈ 2.5x10$^{-3}$ Å/s, ≈ 30 min) were adjusted to grow the desired nanostructure from few MLs to bulky films: dendritic islands of pentacene with thicknesses lower than 10 nm were obtained when P5 was evaporated at 2.5x10$^{-3}$ Å/s for 30 mins on Gr annealed at 150°C. At the same rate, needle-like P5 islands with thicknesses ranging from 15 nm to 30 nm were obtained on Gr annealed at a temperature of 450°C. For bulk films (400 nm), see below, the rate was changed to 0.1 Å/s for 1hr. The substrate is kept at room temperature.

## 4. Macroscopic Gr/P5 device.

The growth of a 400 nm thick P5 film was achieved via the evaporation of P5 through a mechanical mask. After the process described in section 2, we fix, on a sample holder, a mask with openings of 100x100 µm² in front of the sample. The shadow mask is held in an aligned way with the sample that allows the growth of thin films on specific locations on the Gr/SiO$_2$/Si surface (optical image in the inset of Fig. S3).

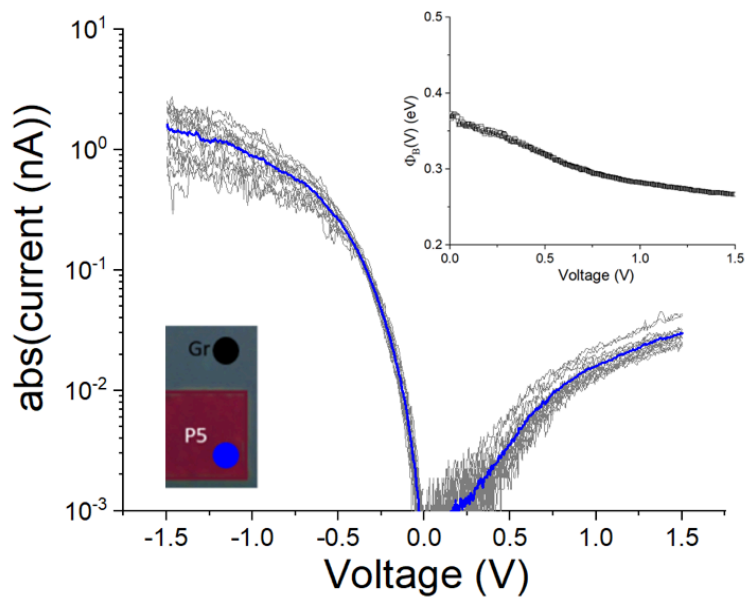

*Figure S3*. I-V data set (gray curves, 16 I-V traces, C-AFM measurements) and mean Ī-V (blue curve) of P5 thin film, 400 nm thick, grown on Gr flake annealed at 150°C. The inset (down left) shows an optical image of the 100 µm x 100 µm P5 film and the underlying Gr flake. Inset (up right) : Voltage dependent energy barrier height for the macroscopic Gr/P5 diode.

The I-V in Fig. S3 cannot be explained by the DSB model, because the high current at V<0 cannot correspond to the saturation current of the P5/PtIr tip diode with a tiny (~15 nm²) surface (even with a very small Schottky barrier height), while the lower current at V>0 corresponds to the saturation current of the large area (100μm x 100μm) Gr/P5 diode, as for the 15-30 nm N-like P5 samples (Figs. 8, S10). Consequently, we assume that only the Gr/P5 diode is measured for this macroscopic sample, the current at V>0 being the forward current of this diode, as also reported for macroscopic P5/Gr diodes (with lithographed top electrodes).[10-13] For the reverse bias (V>0), we considered a modified thermionic emission (MTE) model that takes into account the dependence of the Gr Fermi level with the applied voltage and allow to explain the voltage-driven increase of the saturation current in the reverse regime (V>0).[14, 15] The saturation current $I_S$ is written:

$$I_S = AA^*T^2 e^{-\Phi_B(V)/kT} \quad (S2)$$

with A the contact area, A* the Richardson constant, k the Boltzmann constant, T the temperature and $\Phi_B(V)$ the voltage-dependent Schottky barrier height (SBH) at the Gr/OSC interface. From the mean $\bar{I}$-V, we plot (inset) $\Phi_B(V) = -kT\text{Ln}(I_S/AA^*T^2)$. The zero bias SBH value $\Phi_{B0} \approx 0.38$ eV and the voltage variations of $\Phi_B(V)$ are in agreement with the reported values for macroscopic Gr/P5 diodes (~ 0.25-0.5 eV) with lithographed top electrodes.[10, 11, 16]

5. **C-AFM method**

    *5.1. Discarded I-V traces.*

Some I-V curves were discarded from the analysis:

- I-V traces displaying large and abrupt steps during the scan (contact instabilities).
- At low currents, the I-V traces that reached the sensitivity limit (almost flat I-V traces and noisy I-Vs) and displayed random staircase behavior (due to the sensitivity limit - typically 0.1-1 pA depending on the used gain of the trans-impedance amplifier and the resolution of the ADC (analog-digital converter). Typical examples are shown in Fig. S4.

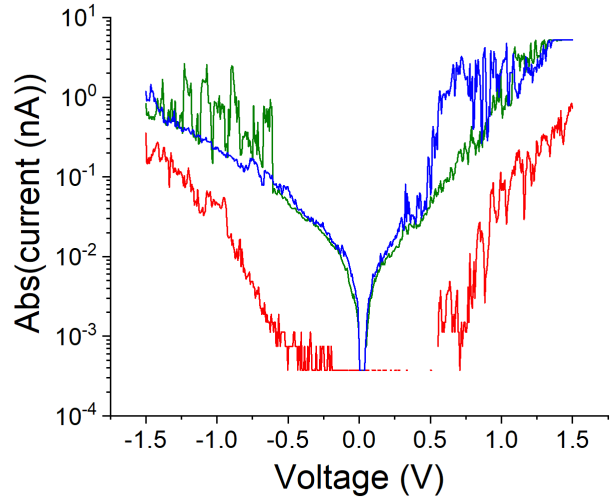

***Figure S4.*** *Three typical I-V traces discarded from the analysis (sample A, 30 nm thick P5 nanostructure).*

5.2. Estimated C-AFM tip contact area and number of contacted molecules.

The loading force was set at ∼ 8 nN for all the I-V measurements, a lower value leading to too many contact instabilities during the I-V measurements. As usually reported in literature[17-20] the contact radius, $r_c$, between the CAFM tip and the P5 surface, and the film elastic deformation, δ, are estimated from a Hertzian model:[21]

$$r_c^2 = \left(\frac{3RF}{4E^*}\right)^{2/3} \quad (S3)$$

$$\delta = \left(\frac{9}{16R}\right)^{1/3}\left(\frac{F}{E^*}\right)^{2/3} \quad (S4)$$

with F the tip loading force (8 nN), R the tip radius (25 nm) and E* the reduced effective Young modulus defined as:

$$E^* = \left( \frac{1}{E^*_{P5}} + \frac{1}{E^*_{tip}} \right)^{-1} = \left( \frac{1-v^2_{P5}}{E_{P5}} + \frac{1-v^2_{tip}}{E_{tip}} \right)^{-1} \tag{S5}$$

In this equation, $E_{P5/tip}$ and $v_{P5/tip}$ are the Young modulus and the Poisson ratio of the P5 and C-AFM tip, respectively. For the Pt/Ir (90%/10%) tip, we have $E_{tip}$ = 204 GPa and $v_{tip}$ = 0.37 using a rule of mixture with the known material data (https://www.webelements.com/). For the P5 nanostructures, we consider the value of an effective Young modulus $E^*_{P5}$ = 15 GPa as measured for thermally evaporated P5.[22, 23] With these parameters, we estimate $r_c$ ≈ 2.2 nm (contact area ≈ 15 nm²) and δ = 0.19 nm.

In a P5 film, the molecules (in the upright position) are organized in stacked monolayers with a herringbone packing in the monolayer and an area per molecule of ≈ 0.25 nm². We estimate that ≈ 60 P5 molecules are connected with the C-AFM tip with our measurement condition.

## 6. Graphene resistance.

The graphene series resistance (graphene in-plane resistance and contact resistance) was measured with the graphene layer connected between two lithographed electrodes separated by a length L = 100μm and width W = 200 μm on a 285 nm thick $SiO_2$ (see section 1 of the SI). Figure S5 shows the typical I-V at low voltages. From the linear behavior, we measured a series resistance of 285 $\Omega$. Rescaled to the typical geometry of the P5 islands measured by C-AFM (Figs. 4 and 5, main text) with a typical length L ∼ 10 μm between the P5 island and the nearest electrode (see Fig. 4-a, main text) and a typical width of the P5 islands of W ∼ 1 μm, the series resistance in the C-AFM geometry is ∼ 5 k$\Omega$.

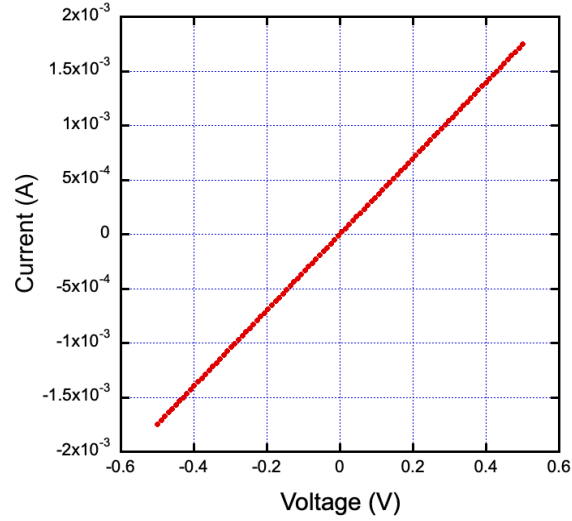

***Figure S5.*** *Current vs. voltage to measure the series resistance (graphene resistance and contact resistance).*

### 7. Details of the models.

***Molecular 2-site model.***

In Eq. (1), the parameters $\varepsilon_{1*}$ and $\varepsilon_{2*}$ are defined by : $\varepsilon_{1*} = (\varepsilon_1 + \varepsilon_2 - \Delta)/2$, $\varepsilon_{2*} = (\varepsilon_1 + \varepsilon_2 + \Delta)/2$ and $\Delta$ is the energy shift between the two orbitals under the application of a voltage V given by $\Delta=(V_i^2+2(\varepsilon_1-\varepsilon_2)V_i+4\tau^2+(\varepsilon_1-\varepsilon_2)^2)^{1/2}$.[24, 25]

In Eq. (2), the internal voltage drop in graphene between the Dirac point and the Fermi energy, $\delta(V)$, is calculated versus the applied voltage V according to Feenstra et al.[26] and given by

$$\delta(V) = \pm\frac{1}{2}\left[\frac{-2C\pi\left(\hbar v_F\right)^2}{e^2} + \sqrt{\left[\frac{2C\pi\left(\hbar v_F\right)^2}{e^2}\right]^2 \pm 4\pi\left(\hbar v_F\right)^2\left(n_0 + \frac{CV}{e}\right)}\right] \quad (S6)$$

with $\hbar$ the reduced Planck constant, $v_F$ the graphene Fermi velocity ($\approx 10^6$ m/s), $n_0$ the intrinsic Gr doping at 0 volt and C the capacitance of the sample (the upper plus sign for $V>-en_0/C$ and the lower minus sign for $V<-en_0/V$), see Fig. S6. We have measured $n_0 = 1.6 \times 10^{12}$ cm$^{-2}$ (see section 2 in this Supporting Information, i.e. at 0 V, $\varepsilon_{DP}$ at 170 meV with respect of Fermi energy) and C =

$\epsilon_{P5}\epsilon_0/d_{P5} \approx 8 \times 10^{-7}$ F/cm² ($\epsilon_{P5} \approx 5$, d ≈ 3 nm for 2 layers of P5, $\epsilon_0$ the vacuum permittivity).

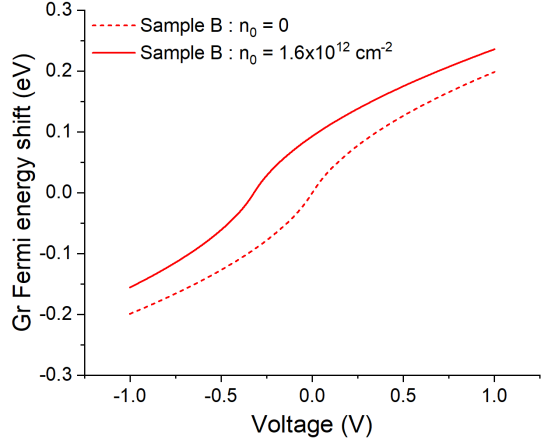

*Figure S6. Typical simulation of Eq. (S6) showing the evolution of the energy difference between the Dirac point and the Fermi level as function of the applied voltage for sample B : no initial doping (dashed line) and with $n_0$=1.6x10¹² cm⁻².*

*Simple fits by parts of the Ī-V curves.*

We consider a modified single energy level (SEL) model to take into account the Gr electrode charging with the applied V as above, substituting for Eqs. 2 and S6 in the following SEL equation:[27, 28]

$$I(V) = N \frac{8e}{h} \frac{\Gamma_1 \Gamma_2}{\Gamma_1 + \Gamma_2} \left[ \arctan\left( \frac{\epsilon_{1,2} + \frac{\Gamma_1}{\Gamma_1+\Gamma_2} eV_i}{\Gamma_1 + \Gamma_2} \right) - \arctan\left( \frac{\epsilon_{1,2} - \frac{\Gamma_2}{\Gamma_1+\Gamma_2} eV_i}{\Gamma_1 + \Gamma_2} \right) \right] \quad (S7)$$

where $\Gamma_1$ and $\Gamma_2$ are now the electronic coupling of the molecule to the two electrodes which are supposed to be different considering the asymmetric geometrical position (Figs. 6d and 6e, main text) of the MO in the junction (while $\Gamma_1 = \Gamma_2 = \Gamma$ in the 2-site model for simplicity).

*Case of the 3L data (point #2).*

We note that the fit of the 2-site model on the mean Ī-V measured at point #2 is a bit worse than at points #1 and #3 (Figs. S8 and S9, $R^2$=0.984 at #2 vs. 0.995 and 0.991 at #3 and #1, respectively, and a larger overlap of the statistical distributions of $\varepsilon_1$ and $\varepsilon_2$) because there is about 3 P5 layers in that case. Nevertheless, the same trends are observed as for the 2L molecular junctions.

### *Double Schottky barrier (DSB) vs. 2-site model.*

Figure S7 shows that the 2-site molecular model (blue line) does not fit well the data for the N-like P5 nanostructure (here the 30 nm thick). Compared to the DSB model (dashed red line, from Fig. S10), the fit is worse and it gives unrealistic energy level values (*e.g.* $\varepsilon_1$=3.1 eV) larger than the HOMO-LUMO band gap of P5 (2.2 eV).

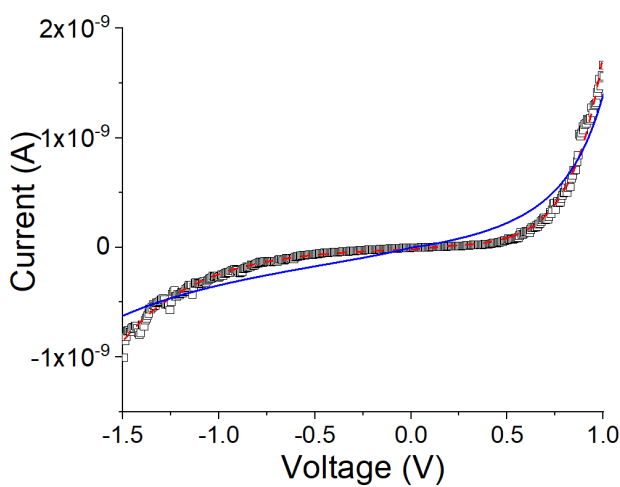

***Figure S7.*** *Fit (blue solid line) of the mean Ī-V curve for the 30 nm thick N-like P5 nanostructure (sample A) with the 2-site model. The fit parameters are $\varepsilon_1$=3.1 eV, $\varepsilon_2$=1.1 eV, Γ=0.38 eV and τ=19 meV. For comparison, the dashed red line is the fit with the DSB model (from Fig. S10).*

We also note that the DSB model is not able to correctly fit the data for the few-layers (2L-3L) Gr/P5 devices (sample B) since it is difficult to consider the existence of a space charge region due to the lack of enough room for band bending in such few-layers devices, albeit several reports pointed out the possibility to use SB model at the nanoscale but with reduced SBH (compared to more bulky devices) because other electron transport mechanisms (*e.g.* tunneling, molecular orbital mediated electron transport) dominate the electron transport at the nanoscale.[29]

## 8. Fits of the datasets of 2-3 MLs P5 devices at points #1 and #2 (sample B).

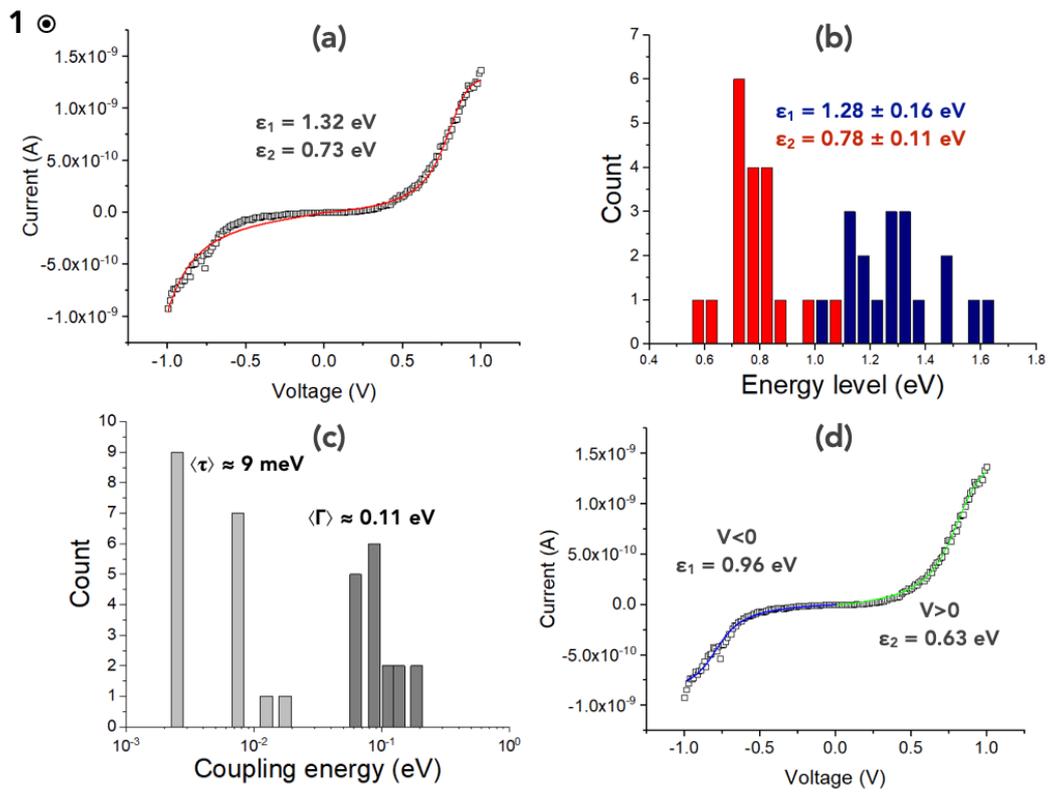

*Figure S8*. Mean $\bar{I}$-V of the dataset measured at point #1 (from Fig. 5, main text), open black squares: (a) fit (red curve) with the 2-site model taking into account the voltage-induced charging of the Gr layer (Eqs. 1, 2, S6); (b) statistical distribution of the energy levels $\varepsilon_1$ and $\varepsilon_2$ (arithmetic mean ± standard deviation); (c) statistical distribution of the coupling energies Γ, τ (arithmetic mean); (d) fit of the mean $\bar{I}$-V with the modified SEL (Eq. S7) for the positive voltage (green curve)

*and negative voltage (blue curve). The fitted energy levels of the molecular orbitals (energy scheme in Fig. 6, main text), $\varepsilon_1$ and $\varepsilon_2$, are given on the figures. The other fit parameters ($\Gamma$, $\tau$, $\Gamma_1$ and $\Gamma_2$) are given in Table S1.*

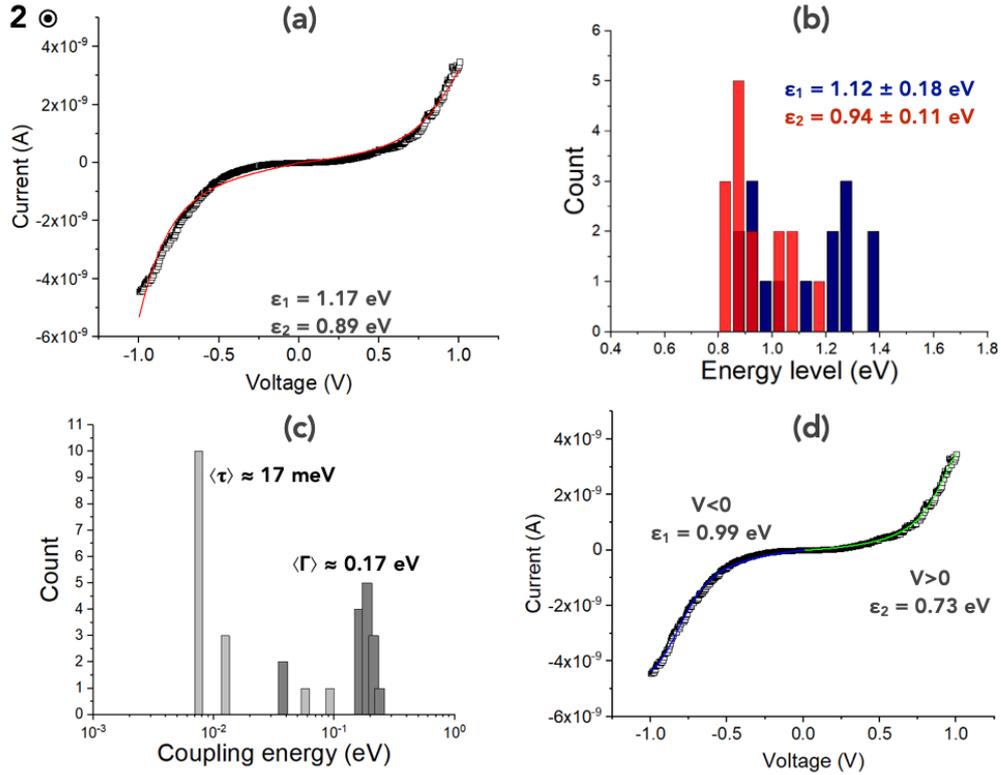

*Figure S9. Mean Ī-V of the dataset measured at point #2 (from Fig. 5, main text), open black squares: (a) fit (red curve) with the 2-site model taking into account the voltage-induced charging of the Gr layer (Eqs. 1, 2, S6); (b) statistical distribution of the energy levels $\varepsilon_1$ and $\varepsilon_2$ (arithmetic mean ± standard deviation); (c) statistical distribution of the coupling energies $\Gamma$, $\tau$ (arithmetic mean); (d) fit of the mean Ī-V with the modified SEL (Eq. S7) for the positive voltage (green curve) and negative voltage (blue curve). The fitted energy levels of the molecular orbitals (energy scheme in Fig. 6, main text), $\varepsilon_1$ and $\varepsilon_2$, are given on the figures. The other fit parameters ($\Gamma$, $\tau$, $\Gamma_1$ and $\Gamma_2$) are given in Table S1.*

|  |  | **2-site model** |  | **SEL (V < 0)** | **SEL (V>0)** |
|---|---|---|---|---|---|
| #1 | Γ (eV) | 0.11 | Γ₁ (eV) | 0.16 | ≈ 0 |
|    | τ (meV) | 9 | Γ₂ (eV) | ≈ 0 | 0.14 |
| #2 | Γ (eV) | 0.17 | Γ₁ (eV) | 0.21 | ≈ 0 |
|    | τ (meV) | 17 | Γ₂ (eV) | ≈ 0 | 0.15 |
| #3 | Γ (eV) | 0.19 | Γ₁ (eV) | 0.22 | ≈ 0 |
|    | τ (meV) | 7 | Γ₂ (eV) | ≈ 0 | 0.18 |

*Table S1*. Values of the fitted electronic coupling parameters (see diagrams in Fig. 10) for the 2-site and SEL models (≈ 0 stands for negligible, i.e. < 0.1 meV). With the SEL model, we obtain $\Gamma_1 > \Gamma_2$ at V < 0 and $\Gamma_2 > \Gamma_1$ at V > 0 as expected from geometrical considerations of the molecular junctions.

## 9. Fits of the datasets of the 20 nm and 30 nm thick P5 devices (sample A).

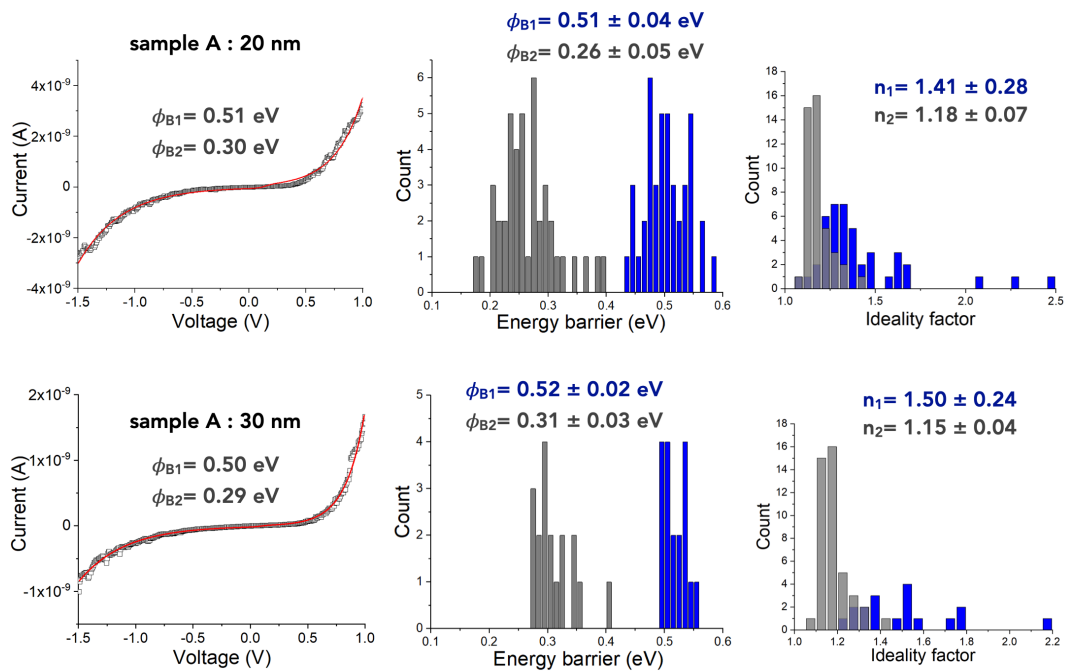

***Figure S10.*** *Fits of the double Schottky diode model on the mean Ī-V dataset of sample A. (a) fit (red line) on the mean Ī-V (open square, data from Fig. 4, main text) of the 20 nm thick N-like P5 island; (b) statistical distribution of the barrier heights by fitting the model on all the individual I-V traces of the dataset shown in Fig. 4 (main text); (c) statistical distribution of the ideality factors. (d-f) Same as (a-b) for the 30 nm thick N-like P5 nanostructures. Note that for these two samples, the fits were limited to 1V because many I-V traces saturate at V > 1V (compliance of the trans-impedance amplifier, see Figs. 4c and 4d in the main text). All the fit parameters are summarized in Table 2 (main text). For the fits, in Eq. S6, the capacitance C is $2.2 \times 10^{-7}$ and $1.5 \times 10^{-7}$ F/cm$^2$ for the 20 and 30 nm thick P5 nanostructures, respectively, and $n_0$ = $1.8 \times 10^{13}$ cm$^{-2}$ (see section 2 in the Supporting Information, i.e. at 0 V, $\varepsilon_{DP}$ at 550 meV with respect of the Fermi level).*